# MuFA (Multi-type Fourier Analyzer): A tool for batch generation of MuMax3 input scripts and multi-type Fourier analysis from micromagnetic simulation output data

Zhiwei Ren, Lichuan Jin, Tianlong Wen, Yulong Liao, Xiaoli Tang, Huaiwu Zhang, Zhiyong Zhong*

*State Key Laboratory of Electronic Thin Films and Integrated Devices, University of Electronic Science and Technology of China, Chengdu, 611731, China*

*Corresponding author at: University of Electronic Science and Technology of China, Chengdu, 611731, China
E-mail address: zzy@uestc.edu.cn

## Abstract

We present a tool for batch generation of input scripts and multi-type Fourier analysis from simulation results for the micromagnetic software MuMax3. The introduction of graphical user interface and parameter-sweeping functionality strongly speed up the input scripts creation and accelerate model optimization processes consequently. Three types of important Fourier analysis methods are provided for the acquisition of the quantitative frequency compositions, the spin-wave dispersion curve and the spatial distribution of spin-wave powers at different frequencies, respectively. Since the Fourier analysis is accelerated by parallel computations, the time cost is reduced to an acceptable level even in the presentation of tens of gigabytes data. With the MuMax3 and our proposal, a complete micromagnetic simulating tool chain from scripts generation to post analysis has been developed.

**Program summary:**
*Program title:* MuFA (Multi-type Fourier Analyzer)
*Licensing provisions:* GNU GPL v3
*Programming language:* Python3
*Computer:* Any computer with Python3 installed
*Operation system:* Any system with Python3 installed
*External routines:* Numpy and Matplotlib modules for Python3

*Nature of problem:*
Magnetization dynamic problems in ferromagnetic elements can be investigated with micromagnetic simulations using MuMax3. However, on one hand, practical modeling processes require a great number of tests with different parameters such as geometry size, which is a heavily time-consuming and error-prone work. On the other hand, a whole set of post Fourier analysis tool for the simulation output data is still missing. Moreover, due to tens of gigabytes data can be generated after a single simulation, the analysis time cost can be greater than the simulation time itself up to several times.

*Solution method:*
The first part of MuFA is a graphical user interface (GUI) tool for the batch generation of input scripts for MuMax3. Numerous input scripts with different geometry size and excitation parameters from each other can be generated at once. The second part of MuFA provides three kinds of important Fourier analysis approaches: frequency components analysis, spatial distribution analysis of spin-wave (SW) powers at different frequencies and the drawing of SW dispersion curve. The file-reading part, which is the most time-consuming part, is accelerated by multiprocessing technique. Therefore, the total time cost is greatly reduced.

*Running time:*
The execution time of scripts generation can be ignored. The time cost of post Fourier analysis strongly depends on the model size and the simulation time period. Two instances used in this paper cost 5 minutes and more than half an hour, respectively.

# 1. Introduction

Novel spintronic devices based on nanoscale ferromagnetic elements have received extensive attention as one of the most promising alternatives to CMOS technology [1-4]. The magnetization dynamics in spintronic systems such as magnetic tunnel junctions (MTJs) [5-7] or magnonic crystals in which spin waves (SWs) act as information carriers [8-10] have been widely investigated. The SWs is one of the most likely candidates of next generation information carriers due to their high operation speed and low power consumption property [11,12].

In order to explore the physical mechanisms under the magnetization dynamic phenomena in these systems, the micromagnetic simulations are generally adopted due to the capacity of handling inhomogeneous complex structures. Owing to its GPU-accelerated (Graphical Processing Units) feature and Go-language [13] based succinct input syntax, MuMax3 has become one of the most popular open-source micromagnetic software with about 800 articles referring to it in just 5 years [14]. However, unlike classic simulator Object Oriented Micromagnetic Framework (OOMMF) [15], a complete micromagnetic simulating tool chain of MuMax3 still has not been developed. On one hand, in practical modeling processes, the user often needs to run a great number of tests with different parameters such as geometry size for the purpose of functionality optimization or investigating the dependencies on specific parameters. It is a heavily time-consuming and error-prone work, although the input scripts are simple. On the other hand, although several post analysis tools for OOMMF output files can be extended to MuMax3 after some modifications [16,17], a whole set of Fourier analysis tool for the studies on SWs is still missing. In addition, since the volume of the output data can be up to tens of gigabytes after a single simulation, using a usual single-process analysis program will result in that its execution time is much greater than the simulation itself.

Here, we present a multifunctional tool: Multi-type Fourier Analyzer (MuFA), which is composed of two independent functional parts, named SBG and MFA which are developed with popular programming language Python [18]. The SBG is a graphical user interface (GUI) tool for the batch generation of input scripts for MuMax3. The MFA is a multiprocessing-accelerated Fourier analysis program for micromagnetic simulation output data. MFA provides three kinds of important Fourier analysis methods: frequency components analysis, spatial distribution analysis of SW powers at different frequencies and the drawing of SW dispersion curve.

We briefly introduce the theoretical background of the micromagnetic simulation in Section 2. Then in Section 3 and 4, we describe the program structure and introduce the running approach respectively in detail. Finally, an example based on a classic magnonic waveguide and performance test on MFA are presented in Section 5.

## 2. Theoretical background

Micromagnetics was introduced firstly by Brown and LaBonte [19]. By treating the ferromagnetic material as a continuous medium, this theory allows one to perform calculations of magnetization behavior at micro- and nanoscale. Since this model looks for the local magnetization configuration satisfying minimum energy principle based on the effective field $H_{\text{eff}}$, the main superiority of micromagnetic approach compared with analytical calculations is the ability of handling inhomogeneous complex structures and non-linear effects. The effective field $H_{\text{eff}}$ can be expressed as following:

$$H_{eff}(r) = H_{ext} + H_{exch} + H_{dem} + H_{anis} , \qquad (1)$$

here, $H_{ext}$, $H_{exch}$, $H_{dem}$ and $H_{anis}$ are the external magnetic field, the exchange field, the demagnetizing field and the anisotropy field, respectively. After the discretization of the hypothetical continuous medium, the $H_{eff}$ in each cell can be expressed as:

$$H_{eff}(r) = \frac{-1}{\mu_0 M_S V_r} \frac{\partial E_r}{\partial m_r} , \qquad (2)$$

where $\mu_0$ is the vacuum permeability, $M_S$ is material saturation magnetization, $V_r$ is the volume of cell $r$, $E_r$ is effective energy in the cell and $m_r$ is the magnetization unit vector of cell $r$. When the magnetization vector is disturbed and moved from its equilibrium position, a precession with damping takes place around the effective field. This dynamic process is described by Landau–Lifshitz–Gilbert (LLG) equation, which is an ordinary differential equation in time domain [20,21]:

$$\frac{d\vec{M}}{dt} = -|\gamma|\vec{M} \times \vec{H}_{eff} + \frac{\alpha}{M_S} \vec{M} \times \frac{d\vec{M}}{dt} , \qquad (3)$$

where $\gamma$ is the gyromagnetic ratio and $\alpha$ is the dimensionless Gilbert damping factor. In order to describe spin transfer torque (STT) effect, LLG equation can be extended by adding two STT terms [22,23]:

$$\frac{d\vec{M}}{dt} = -|\gamma|\vec{M} \times \vec{H}_{eff} + \frac{\alpha}{M_S} \vec{M} \times \frac{d\vec{M}}{dt} + |\gamma|\beta\varepsilon\vec{m} \times (\vec{m}_p \times \vec{m}) - |\gamma|\beta\varepsilon'\vec{m} \times \vec{m}_p , \qquad (4)$$

$$\beta = \frac{\hbar J}{\mu_0 |e| M_S t} , \qquad (5)$$

$$\varepsilon = \frac{\vec{P}\Lambda^2}{(\Lambda^2+1)+(\Lambda^2-1)(\vec{m}\cdot\vec{m}_p)} \ , \tag{6}$$

here, $m$ and $m_p$ are magnetization unit vector and spin polarization unit vector respectively. $J$ is current density, $t$ is the thickness of the free layer, $\vec{P}$ is spin polarization, $\Lambda$ is a parameter characterizes the spacer layer and $\varepsilon'$ is the phenomenological secondary spin-torque coefficient.

Due to the computation for the total energy of a magnet depends on the spatial magnetization distribution, LLG equation is too complicated to be solved analytically in most instances. Therefore, a numerical method must be utilized, which is usually realized by the introduction of a finite element method (FEM) [24-26] or finite difference method (FDM) [14,15,27]. Owing to the use of tetragonal mesh cell, the geometrical flexibility of FEM is higher compared with FDM in which cuboid cell is applied. However, the cost of high flexibility is the consumption of additional computing resources and consequently, the limitation of the size of the model. Thus, currently two of the most popular open-source micromagnetic software — OOMMF and MuMax3 — are both based on FDM.

## 3. Structure of the program

MuFA consists of two independent functional modules — SBG and MFA. SBG is a GUI program that is responsible for the batch generation of input scripts for micromagnetic simulator MuMax3. MFA is a program that is responsible for the Fourier analysis from the micromagnetic simulation results. Results from all of OOMMF-like micromagnetic simulator, e.g. MuMax3, can be analyzed with MFA.

### *3.1 SBG*

SBG consists of a main window with several buttons that will direct users to subwindows which allowing for the generation of MuMax3 simulation codes (For a detailed description of MuMax3 Application Programming Interface (API), see: http://mumax.github.io/api.html). For the reason that the classification of the simulation elements in original API is too scattered, we integrate and reclassify the simulation elements according to the practical modeling process (as shown in Fig. 1). For understanding and better utilizing the program, user should refer to the API linked above firstly. Current version of SBG is developed in accordance with the version 3.9.1 of MuMax3. Since MuMax3 is an open-source program and is continually updated, maybe there will be several new properties which are not included in SBG in newer version of MuMax3. User can introduce these properties manually into the edit window of the main window. Below, we introduce the functionality of each button and entry of SBG briefly. The detailed instruction can be found in the README file enclosed to the distributed file.

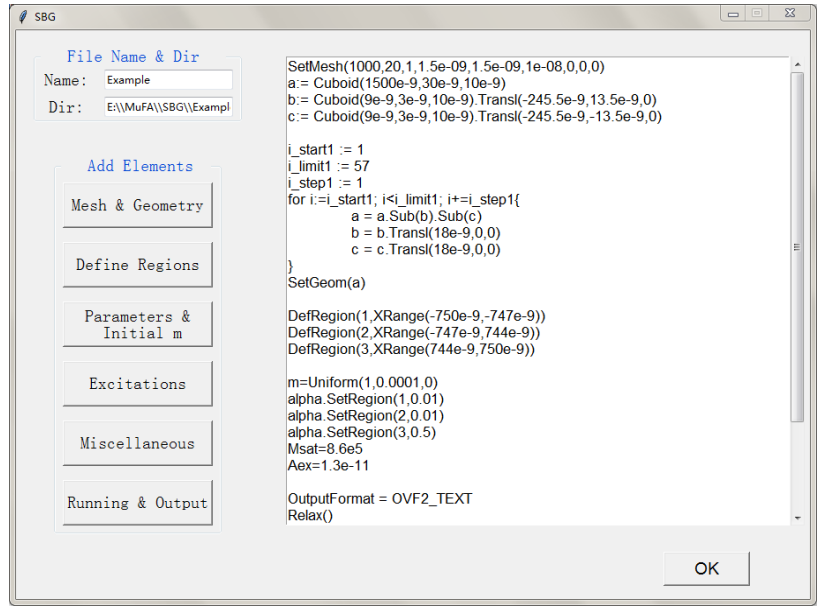

Fig. 1. The GUI of SBG main window.

*3.1.1 File Name & Dir*
• 'Name': Input the basic file name of the scripts into this entry.
• 'Dir': Input the directory in which user wants to generate the scripts into this entry.

*3.1.2 Mesh & Geometry*
　'Mesh & Geometry' button is responsible for the definitions of the mesh size used in simulations and sample geometry. As shown in Fig. 2, there are two modes for mesh definition. By clicking the tabs, user can switch the definition modes:
• 'Fixed mode': user inputs a fixed mesh size and the mesh sizes are the same in all of scripts.
• 'Sweep mode': user inputs a start size, an end size and a step size for x, y, z direction respectively. This leads to the mesh sizes in output scripts change step by step from each other.
'cell' series entries define the mesh cell size of the simulation and 'PBC' series entries define the periodic boundary conditions. After the definition of all mesh parameters, click 'Add This Mesh' button to input the mesh term into the edit window.
　To define the geometry of the sample, user selects several shapes, such as cuboid and the transformation for the shapes from the 'Shape' menu and 'Method' menu. 'Loop' menu provides three kinds of loop template in Go language. It is utilized to generate complicated periodic geometry, e.g. magnonic crystal. Click 'Add shape' button to add current shape to the final geometry. After the definition of the geometry, click 'Add This Geom' button to input the geometry term into the edit window.
　Usually, the parameters of the geometry are directly keyed in the edit window. If 'Sweep mode' is used during the definition of the mesh, user can utilize arbitrary number of "x", "y" or "z" instead of real numbers to change the geometry size in order to make it be consistent with the mesh size. This is not only effective in geometry definition, but also valid in other definitions involved regions. The following three buttons work the same in all of subwindows.
- 'Undo': it is responsible for the undo operation.
- 'Cancel & Quit': delete all of content in edit window and kill this subwindow.
- 'Confirm & Quit': add all of content to the edit window of the main window and kill this subwindow.

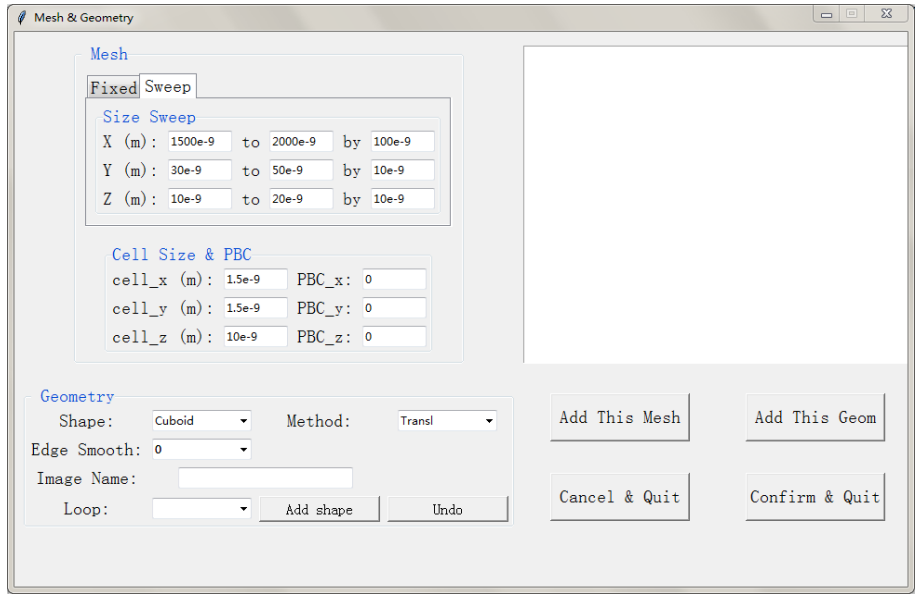

Fig. 2. The GUI of SBG 'Mesh and Geometry' subwindow.

*3.1.3 Define Regions*

'Define Regions' button is responsible for the definition of the simulation space division. The definition of the regions is similar to the process of geometry definition (Section 3.1.2). User selects the 'Shape' and the 'Method', then clicks 'Add Shape' button to add current shape to the final region shape. Finally, click 'Add This Region' button to add this region to the edit window. Regions are automatically numbered. In other definition procedures that involved in setting a parameter or excitation to specified regions, the regions are automatically numbered too. Parameters are directly input into the edit window. 'Delete Region' button is responsible for the undo operation.

*3.1.4 Parameters & Initial m*

'Parameters & Initial m' button is responsible for the definitions of material parameters and the initial magnetization configuration.
- 'Initial m': initial magnetization configuration is determined via this menu, and additional operation can be added from 'Method' menu. After the definition, click 'Add Initial m' button to add the initial magnetization configuration to the edit window.
- 'Parameters': all of material parameters are specified by this menu. First, choose the material parameter, then key the corresponding value into the 'Function' entry. Next, add additional operations from 'Method' menu. Finally, click 'Add This Para' button to add this parameter to the edit window.

*3.1.5 Excitations*

'Excitations' button is responsible for the definition of magnetic field or charge current excitations used in the simulation. There is no difference between the definitions of these two kinds of excitations except for the unit. So we only introduce the definition of the magnetic field excitations.

As depicted in Fig. 3, due to the excitation field is a vector, user should input the 'Function' of the excitation in the form of "vector(X, Y, Z)". It should be noted that the amplitude and the frequency values must be replaced by "amp" and "f". Specified values are input into the entries at the left bottom of the window. The 'Method' menu provides available methods corresponding to the excitation. Similar to the mesh size and the geometry, the amplitude and frequency of excitations are also parameters which can be swept. The detailed process of the sweep definition can be found in Section 3.1.2.

'Misc' menu provides several seldom used elements. Click 'Add This Misc' button to add one. 'Add This B_ext' and 'Add This J_ext' button are responsible for adding magnetic field excitation term and current excitation term to the edit window, respectively.

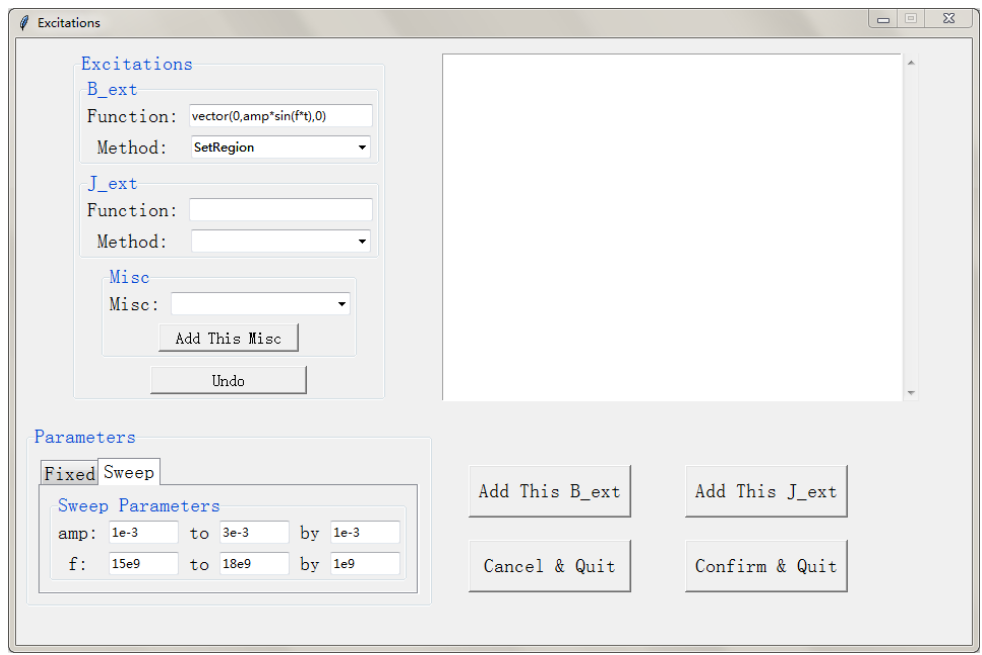

Fig. 3. The GUI of SBG 'Excitation' subwidow.

*3.1.6 Miscellaneous*

'Miscellaneous' button is responsible for adding several special elements to the script. There are four categories in the miscellaneous subwindow. They are 'Moving Simulation Window', 'Magnetic Force Microscopy', 'Extensions' and 'For loops'. For the first three categories, there are detailed descriptions in http://mumax.github.io/api.html. The 'For loops' menu offers the template of loop in

Go language from one cycle to triple cycle. We introduce this class to increase the flexibility of the modeling procedure. There are four buttons corresponding to the four categories. Click corresponding button to add the selected item to the edit window.

*3.1.7 Running & Output*

'Running & Output' button is responsible for the definition of the scheduling output quantities and running type.
• 'Export Format': provides options that determine the coding form of the simulation output files. If user want to apply MFA for the analysis subsequently, he/she should choose "OVF2_TEXT" option.
• 'Output Type': specifies the Output type.
• 'Slicing': For the purpose of saving storage space, user can only save part of the output via selecting one of the items provided by this menu.
• 'Quantities': specifies output quantity.
• 'Method': provides corresponding operations for the output quantity.
• 'Running': provides options of running type and parameters of the differential equation solver.

*3.1.8 OK*

After finishing the input of all of essential elements mentioned above, click 'OK' button to generate scripts in the specified directory.

*3.1.9 Other features*

The name of the generated files contain the information of model size and excitations applied in the simulation. CTRL + C and CTRL + V shortcuts are available in all edit windows.

*3.2 MFA*

MFA is composed of two categories of modules. The first one reads data from a set of vector field files and reorganizes them into a single matrix. The second one performs further processes on the matrix obtained from the first part, and computes different types of Fourier transform in terms of the instruction from the user. These two processes are automatically connected.

*3.2.1 Reorganization of data*

The vector field output files (.ovf) used in Fourier analysis specify magnetizations as a function of spatial position. Magnetization vectors are ordered with the index in the x direction incremented first, then the index in the y direction, and the index in the z direction last. Each magnetization vector is represented by its three components on x, y and z axis. Therefore, there are three columns of data in each vector field file, but only one of them is selected and read into the memory. This can evidently reduce the time cost and the demand of the memory capacity in comparison to reading all of data contained by the file. Then the data are reorganized into a single matrix that contains the information of magnetization temporal evolution as shown in Fig. 4. The variables $t'_{max}$, $x'_{max}$, $y'_{max}$ and $z'_{max}$ represent the desired maximum time point and space point involved in further calculations. Although only the maximum values are depicted in the figure, the minimum values can also be specified. Thus, both the simulation time span and the simulation space area used in Fourier analysis can be arbitrarily determined in terms of instructions.

However, since GPU-accelerated micromagnetic simulators have been widely used, tens of gigabytes data can be generated after a single simulation. The trick mentioned above alone cannot decrease the total time cost effectively. The execution time of the program is still much greater than that of the simulation itself. Hence, we apply the multiprocessing technique to accelerate this process. Vector field files are divided into several parts according to the number of the assigned processes and each process reads one part of the files only. Then reorganization is performed to merge the data into a matrix for the next step process.

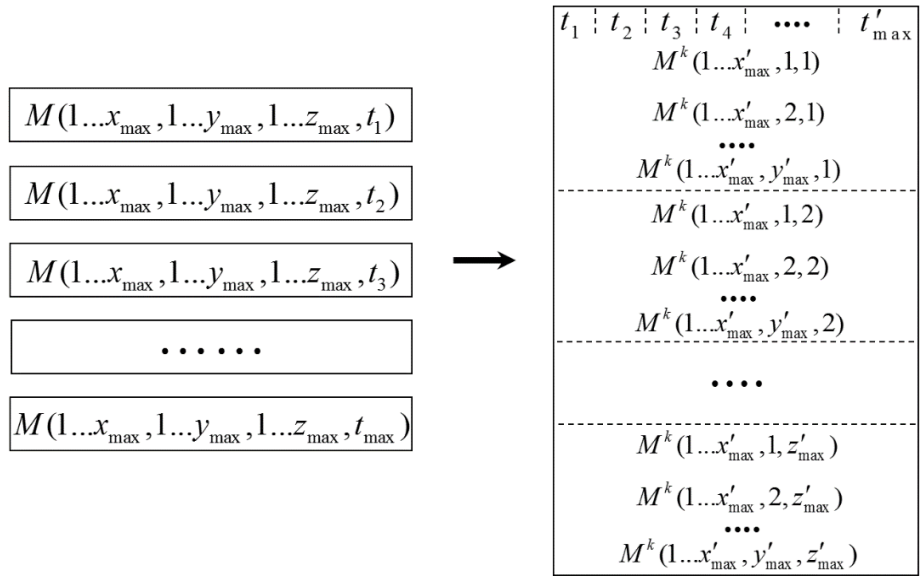

Fig. 4. Schematic of the data reorganization. The arrow indicates the direction of the data flow. A solid line box reprents a file (left side) or a matrix (right side). The variables $t_{max}$, $x_{max}$, $y_{max}$ and $z_{max}$ represent the maximum time point of the simulation and the maximum space point of the model. The variables $t'_{max}$, $x'_{max}$, $y'_{max}$ and $z'_{max}$ represent the desired maximum time point and the space point. Each row of the matrix contains the information of the magnetization temporal evolution of a single simulation cell.

*3.2.2 Computations of Fourier transform*

MFA can provide one-dimensional and two-dimensional Fourier analysis. For one-dimensional Fourier analysis, fast Fourier transform (FFT) is performed on each column of the matrix acquired from previous stage, i.e. FFT in the time domain. Then we take the average on the absolute value summation of FFT results and compute corresponding frequency points. MFA provides two types of one-dimensional Fourier analysis. One of them is "FFT" that shows normalized frequency compositions of the magnetization dynamics, e.g. SWs. The other one called "Spectrum" that produces a colomap to visualize the spatial distribution of SW powers at different frequency along the wave-vector direction. The two-dimensional Fourier analysis on SWs is called "Dispersion curve". It is one of the most important analysis approaches in the magnonic researches. The computation procedure is presented in Fig. 5 [28]. By taking the average on the cross section that perpendicular to the wave vector, the matrix is compressed to a one-dimensional vector. Then two-dimensional FFT is applied to transform it from the time-space domain to the frequency and wave-vector domain. Finally, the result is visualized by a colormap. According to previous study on the SW dispersion curve computations [28], the introduction of a Chebyshev window function in the calculation process can effectively reduce the spectral leakage and the scalloping loss. Therefore, a two-dimensional Chebyshev window function is optional in MFA when a dispersion curve calculation is performed. It should be noted that for the "Spectrum" and "Dispersion curve" analysis, the wave-vector direction have to lie in the x axis. If wave vectors in other directions are desired, the simulation model should be rotated to meet the condition mentioned above. It is can be easily realized in both MuMax3 and OOMMF.

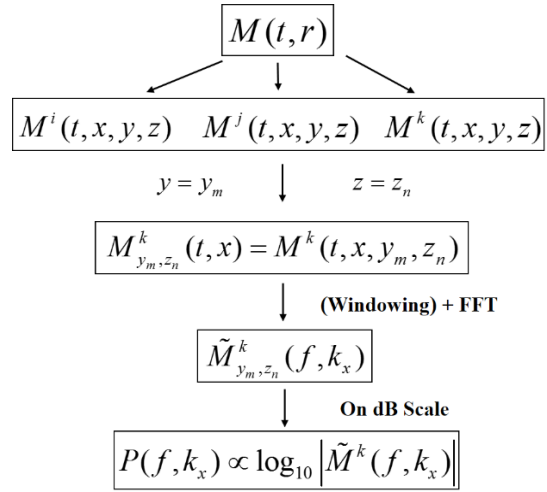

Fig. 5. Steps involved in obtaining a spin-wave dispersion curve.

## 4. Running

In order to run both SBG and MFA, Python3 interpreter should have been installed and for MFA, additional modules Numpy [29] and Matplotlib [30] are required. If user wants to utilize window function in dispersion curve calculation, a more powerful Python-based ecosystem Scipy [31] is required. This is not indispensable.

To run SBG, the only thing to do is executing the file 'SBG.py' by Python3 interpreter. For MFA, we present a job-script controlled version that is accelerated by adopting multiprocessing technology (MFA_MultiCores.py), and a version using GUI (MFA_GUI.py). Unfortunately, limited by the technique, current GUI version cannot utilize the multiprocessing acceleration. Therefore, it is only suitable to perform analysis on the results from small-size and short-time simulations.

## 5. Example

As an example, we perform an analysis on a classic magnonic waveguide model proposed by Lee *et al* [32] using MFA. The simulation script is generated by SBG and the simulation results are from MuMax3. In order to avoid making the article too lengthy, the specific procedure of the simulation scripts generation is omitted. It can be found in the example README file, in which we present an instance of how to generate 90 script files with different geometry sizes and excitations at once. The simulation input script used here is "e_1000_20_1_1.0e+00_1.0e+11.txt" that can be found in the example output folder of SBG. Below, we give a description of this magnonic waveguide model and show the analysis results from MFA.

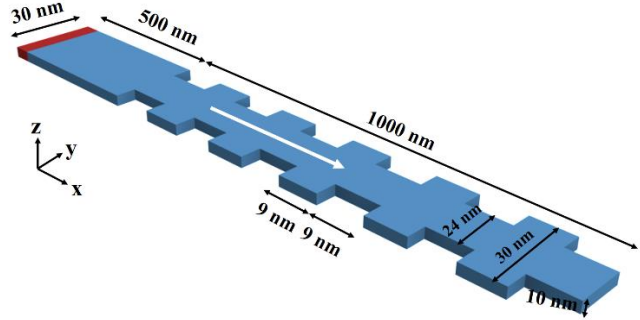

Fig. 6. Schematic of the magnonic crystal waveguide used as the example. The direction of the initial magnetization is +x, as indicated by the white arrow. In order to generate spin waves propagating in the +x direction, A sinc function field with $b_0$ = 1.0 T is applied along the y direction only to the local area of 1.5×30×10 nm³, indicated by red color.

The schematic of the magnonic waveguide is shown in Fig. 6. The material used here is Permalloy (Py), the chosen material parameters are as follows [32]: the saturation magnetization $M_s$ = 8.6×10⁵ A/m, exchange constant $A_{ex}$ = 1.3×10⁻¹¹ J/m, Gilbert damping = 0.01. The damping at the end of the magnonic waveguide is set to be 50 times that of other parts to prevent SWs reflection. The cell size is set to 1.5×1.5×10 nm³. To excite spin waves with rich frequency components, a "sine cardinal (sinc)" function field $b_y(t) = b_0 \sin[2\pi f_c(t + t_0)] / 2\pi f_c(t + t_0)$ with $b_0$ = 1.0 T, $f_c$ = 100 GHz and $t_0$ = 0.1 ps is applied along the y direction only to the local area of 1.5×30×10 nm³, indicated by red color. Due to the coupling between the lowest spin-wave mode and other spin-wave modes caused by the periodical width-modulation, spin waves propagating in this magnonic waveguide will show complex band structure and several wide band gaps [32].

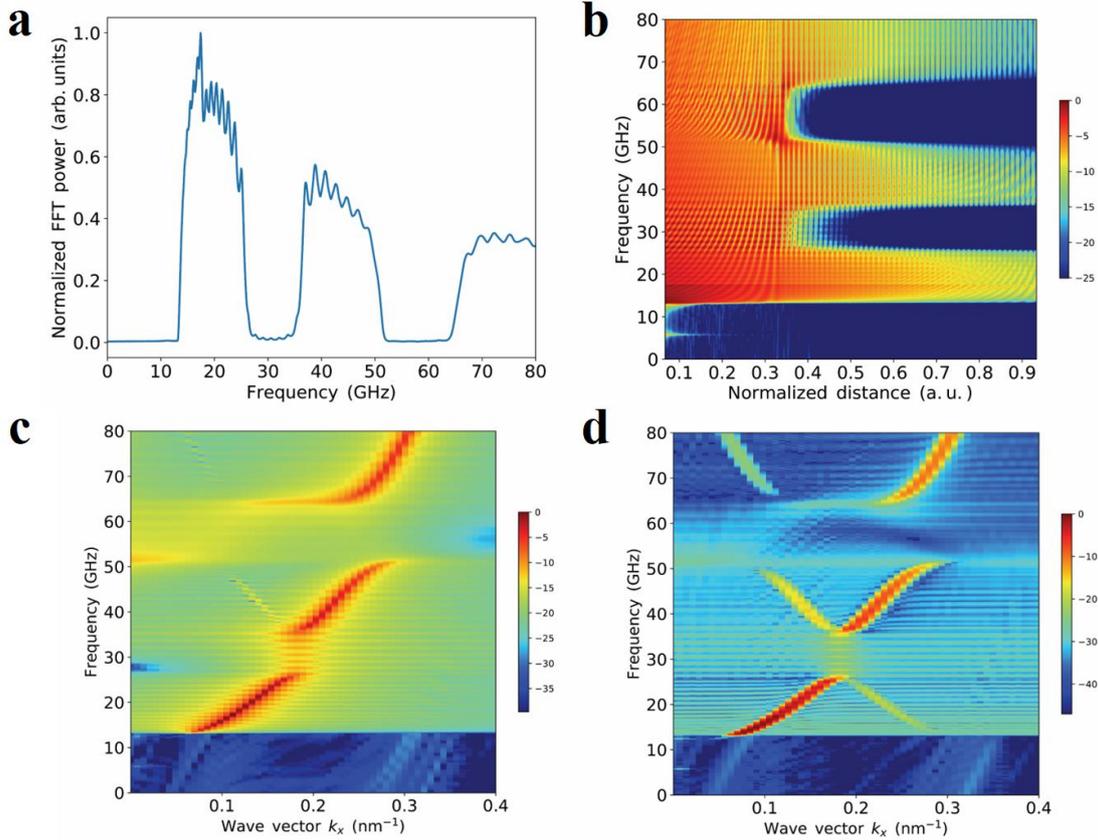

Fig. 7. Results of different type of Fourier analysis computed by MFA. Sampling range in the y-direction is from 13.5 to 16.5 nm. (a) The spectrum that indicates the frequency compositions of the spin wave propagating in the magnonic crystal (FFT). Sampling range in the x-direction is from 500 to 1400 nm. (b) The colormap that shows the spatial distribution of the spin-wave power at different frequencies along the direction of the wave vector (Spectrum). Sampling range in the x-direction is from 100 to 1400 nm. (c) The spin-wave dispersion curve without a window function (Dispersion curve). Sampling range in the x-direction is from 600 to 1400 nm. (d) The spin-wave dispersion curve with a two-dimensional Chebyshev window function (Dispersion curve). The sampling area is the same as that in (c).

Fig. 7 shows the analysis results computed by MFA. As indicated in Fig. 7(a), the spectrum quantitatively describes the normalized frequency components of the SWs propagating in the magnonic waveguide. It can be found that there are three obvious forbidden bands. The first one is the intrinsic potential barrier that is below 14 GHz, the second one is between 26 and 36 GHz, and the last one is between 50 and 66 GHz, whereas the locations and the shapes of these forbidden bands can be visualized by the "Spectrum" analysis provided by MFA as shown in Fig. 7(b). Since the last two forbidden bands only lie in the width-modulated area, it can be assumed that their origins are related to the width modulation. For the purpose of revealing the physical origin of the forbidden bands, the dispersion curve must be calculated. As depicted in Figs. 7(c) and (d), the relationship between the frequency and the wave vector has been obtained. It can be found that once the window function is applied, the power distribution becomes more concentrated in comparison to the case without window function. Therefore, MFA provides a set of powerful analysis methods which are favorable to the SW research. All of the results are consistent with previous study [32].

The analysis time of the example was about 5 minutes on our computer. To validate the performance of MFA in the presentation of large-scale data, an additional simulation of which the number of the output files was 10,000 and the total data size was 32.1 GB has been performed. The test was run on a personal computer that contains an Intel i5-4460 4-core processer with frequency of 3.2 GHz, and a NVIDIA GeForce GTX 750 graphics card. The simulation time cost was 34 minutes, whereas the execution time of MFA was 32-38 minutes for all types of Fourier analysis. These two time costs were in the same level in this case. Although the simulation time can be further decreased with more powerful GPU, this time cost of the analysis is still acceptable, not to mention that the application of the processor with more cores also can improve the performance of MFA. We also performed a single-process analysis using the GUI version MFA. The execution time was 100-108 minutes which was about three times of the 4-processes case.

## 6. Summary

We have developed an open-source tool for assisting users with both the batch generation of input scripts for micromagnetic simulator MuMax3 and multi-type Fourier analysis from micromagnetic simulation results. Owing to its GUI and parameter sweeping functionality, the first part of the program can strongly speed up the scripts creation and accelerate model optimization processes, consequently. The second part of the program provides three different types of Fourier analysis methods for the acquisition of the quantitative frequency compositions, the dispersion curve and the spatial distribution of spin-wave powers at different frequencies, respectively. The analysis program is accelerated by multiprocessing technique. Thus, the analysis time cost is reduced to an acceptable level even in the presentation of tens of gigabytes data. With the MuMax3 and our proposal, a complete micromagnetic simulating tool chain from scripts generation to post analysis has been developed.

## Acknowledgments

This paper is supported by the National Natural Science Foundation of China under grant Nos. 61734002, 61571079 and 51702042, the National Key Research and Development Plan (No. 2016YFA0300801); and the Sichuan Science and Technology Support Project (No. 2017JY0002).

## Additional information

**Declarations of interest**: None
No figure in print should use color.